\documentclass[11pt,twoside]{article}
\usepackage{graphicx,epsfig,natbib,epstopdf}
\usepackage{CS18}

\begin{document}

\title{Bridging the gap on tight separation brown dwarf binaries}

\author{Daniella C. Bardalez Gagliuffi$^{1,2}$, Adam J. Burgasser$^{1}$, Christopher R. Gelino$^{2}$, Carl Melis$^{1}$, Cullen Blake$^{3}$}

\affil{$^1$Center for Astrophysics and Space Sciences, University of California, San Diego. 9500 Gilman Dr., Mail Code 0424, La Jolla, CA 92093, USA.}
\affil{$^2$NASA Exoplanet Science Institute, Mail Code 100-22, California Institute of Technology, 770 South Wilson Ave, Pasadena, CA 91125, USA.}
\affil{$^3$Department of Physics and Astronomy, University of Pennsylvania, 219 S. 33rd St., Philadelphia, PA 19104, USA.}

\begin{abstract}
Multiplicity is a key statistic for understanding the formation of very low mass (VLM) stars and brown dwarfs. Currently, the separation distribution of VLM binaries remains poorly constrained at small separations ($< 1$ AU), leading to uncertainty in the overall binary fraction.  We approach this problem by searching for late M/early L plus T dwarf spectral binaries whose combined light spectra exhibit distinct peculiarities, making their identification independent of separation. We define a set of spectral indices designed to identify these systems, and use a spectral template fitting method to confirm and characterize spectral binary (SB) candidates from a library of 738 spectra from the SpeX Prism Spectral Libraries.  We present twelve new binary candidates, confirm two previously reported candidates and rule out other two previously reported candidates. All of our candidates have primary and secondary spectral types between M7$-$L7 and L8$-$T8 respectively. We find that blue L dwarfs and subdwarfs are contaminants in our sample and propose a method for segregating these sources. If confirmed by follow-up observations, these systems may potentially add to the growing list of tight separation binaries, giving further insight into brown dwarf formation scenarios.
\end{abstract}

\section{Separation Distribution of Very Low Mass (VLM) binaries}

Whether brown dwarfs form as stars from bottom-up accretion or as planets from a pre-stellar disk is still a matter of debate. The necessary conditions for the formation of brown dwarfs require high density regions and low Jeans' mass in the natal gas, as well as limited mass accretion after the onset of collapse. Several theories have been proposed to explain these scenarios. These include ejection of pre-stellar cores~\citep{rei01}, photo erosion of pre-stellar cores~\citep{whi04}, disk fragmentation~\citep{sta09} and turbulent fragmentation~\citep{pad02}. A good way of determining the relative occurrence of these formation theories is by studying multiplicity.

\begin{figure}
\centering
\includegraphics[scale=0.4]{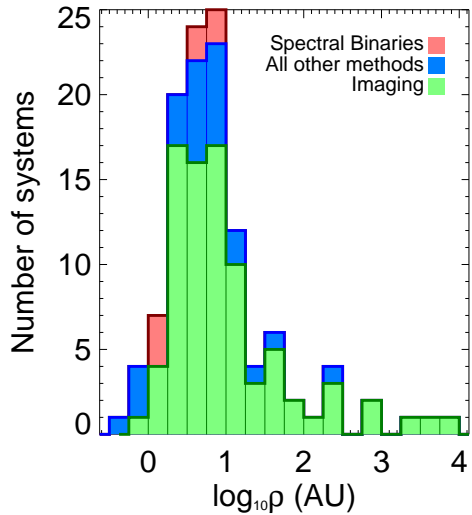}
\caption{Separation distribution of confirmed VLM binaries. The sample includes 115 systems of which 84 were discovered by imaging (green), 7 were spectral binaries before confirmation (red) and the rest were discovered by microlensing, radial velocity or astrometric variability (blue). Figure from Bardalez Gagliuffi et al. (2014, submitted).}\label{fig:sepdist}
\end{figure}

The shape of the separation, mass ratio and eccentricity distributions is a direct consequence of the mechanisms that resulted in brown dwarf formation~\citep{bat09}. While direct imaging has been the most prolific observational method to identify binaries (73\% of confirmed systems~\footnote{VLMbinaries.org}), its inherent limit in spatial resolution suggests incompleteness. For a typical field brown dwarf at a distance of $20-40$pc, the best angular resolution that can be achieved with the Hubble Space Telescope (HST) or the Keck II Telescope using adaptive optics (AO) is $0\farcs2$, leading to a projected separation of $4-8$AU, which brackets the peak of the observed distribution (Figure~\ref{fig:sepdist}). It is possible that this concurrence is a consequence of the limit in imaging resolution rather than a real peak. 

\section{Spectral Binaries}

An alternative, separation-independent method to identify tightly-bound binaries is to look at their unresolved, blended-light spectra. We will use the term \emph{spectral binaries} for those objects whose combined-light spectrum shows distinct peculiarities that arise when the highly structured NIR spectra of single M, L and T dwarfs are combined together. In this study we adapt the technique of~\citet{bur10} to search for spectral binaries composed of late-M or early-L primaries and T dwarf secondaries. Since M dwarfs are the most common stars in the galaxy as well as the brightest VLM objects, and because their spectra are decidedly different from the spectral of T dwarfs, binaries composed of these spectral types are excellent candidates for identification as spectral binaries. In addition, binaries with a late-M/early-L primary and a T dwarf secondary can straddle the Hydrogen burning limit, making them interesting probes for brown dwarf thermal evolution.

\begin{figure}
\centering
\includegraphics[scale=0.4]{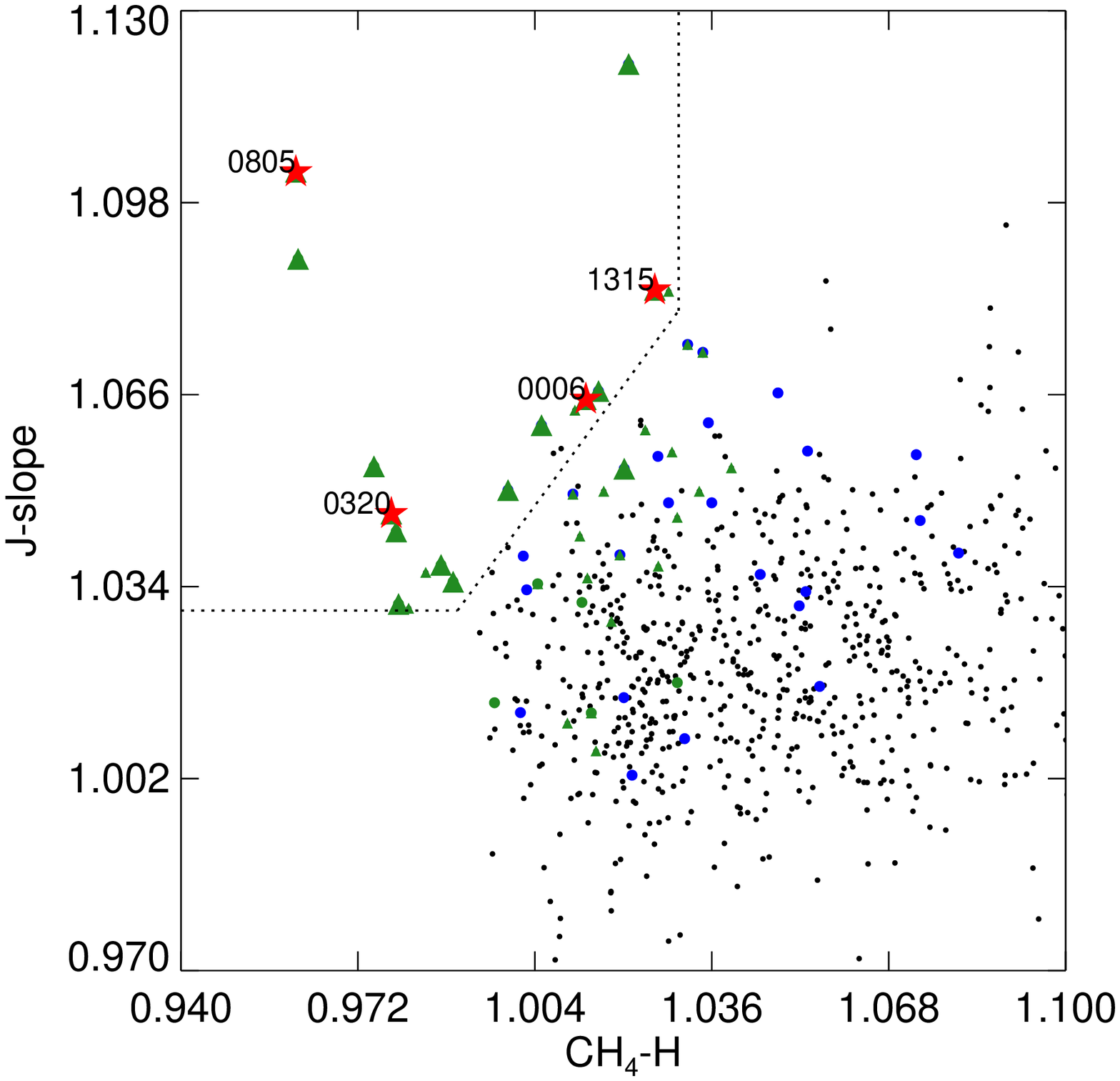}
\includegraphics[scale=0.4]{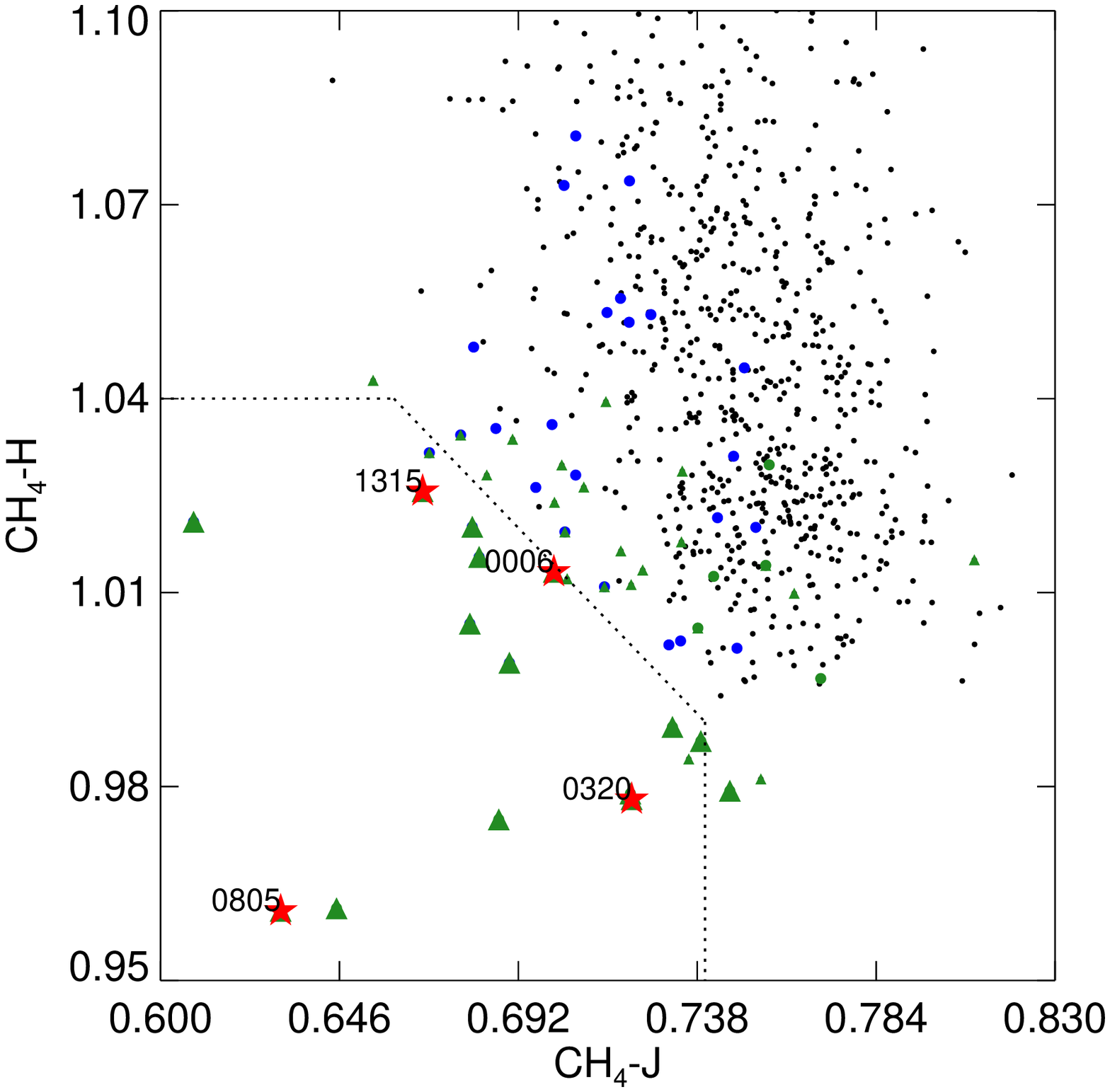}
\caption{Comparison of spectral indices. The red labelled stars indicate the four benchmark binaries, the green triangles represent index-selected candidates, the blue circles are unusually blue objects as noted in the literature. Figure from Bardalez Gagliuffi et al (2014, submitted).}\label{fig:indices}
\end{figure}

\begin{figure}
\centering
\includegraphics[scale=0.3]{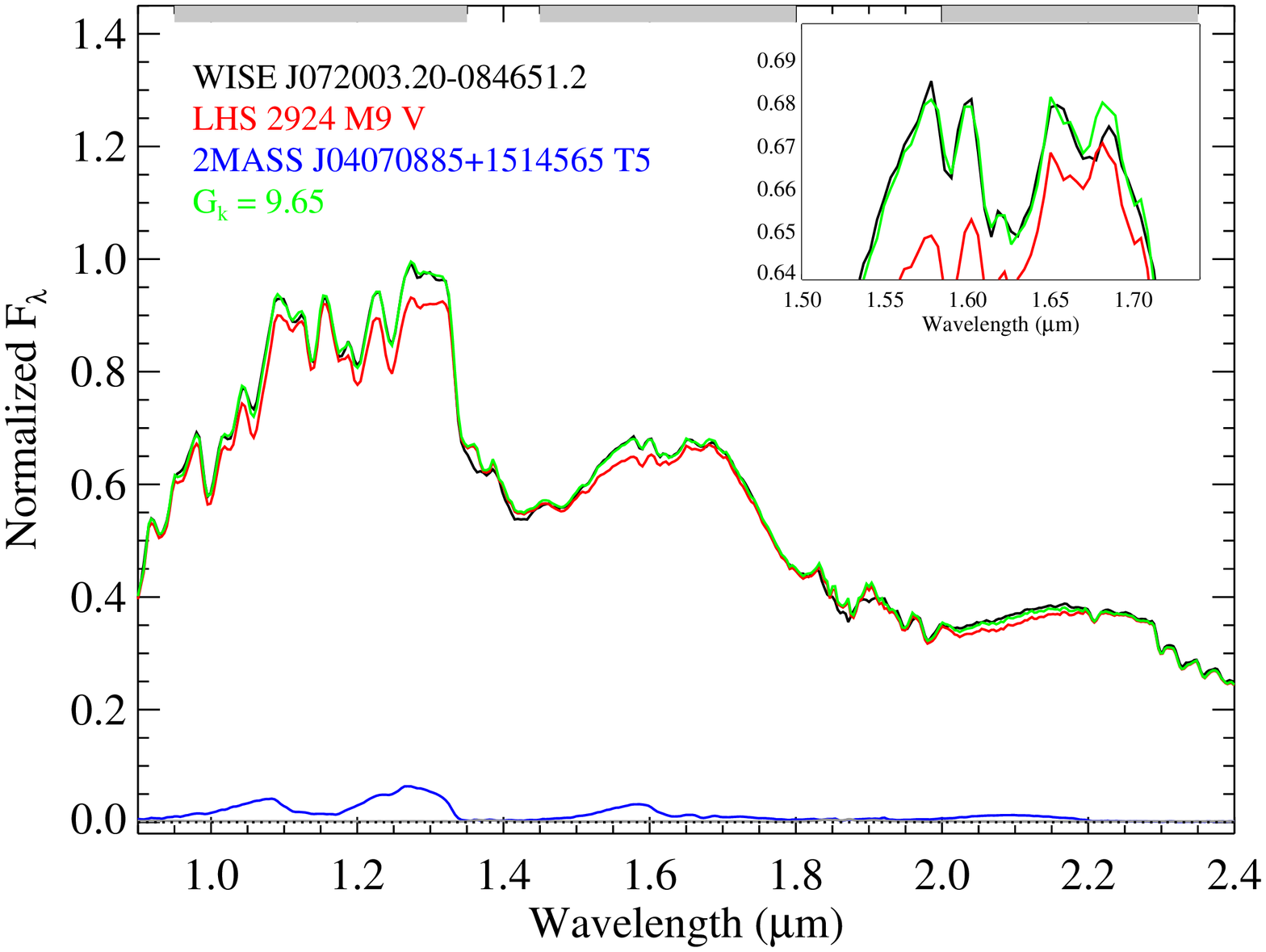}
\includegraphics[scale=0.3]{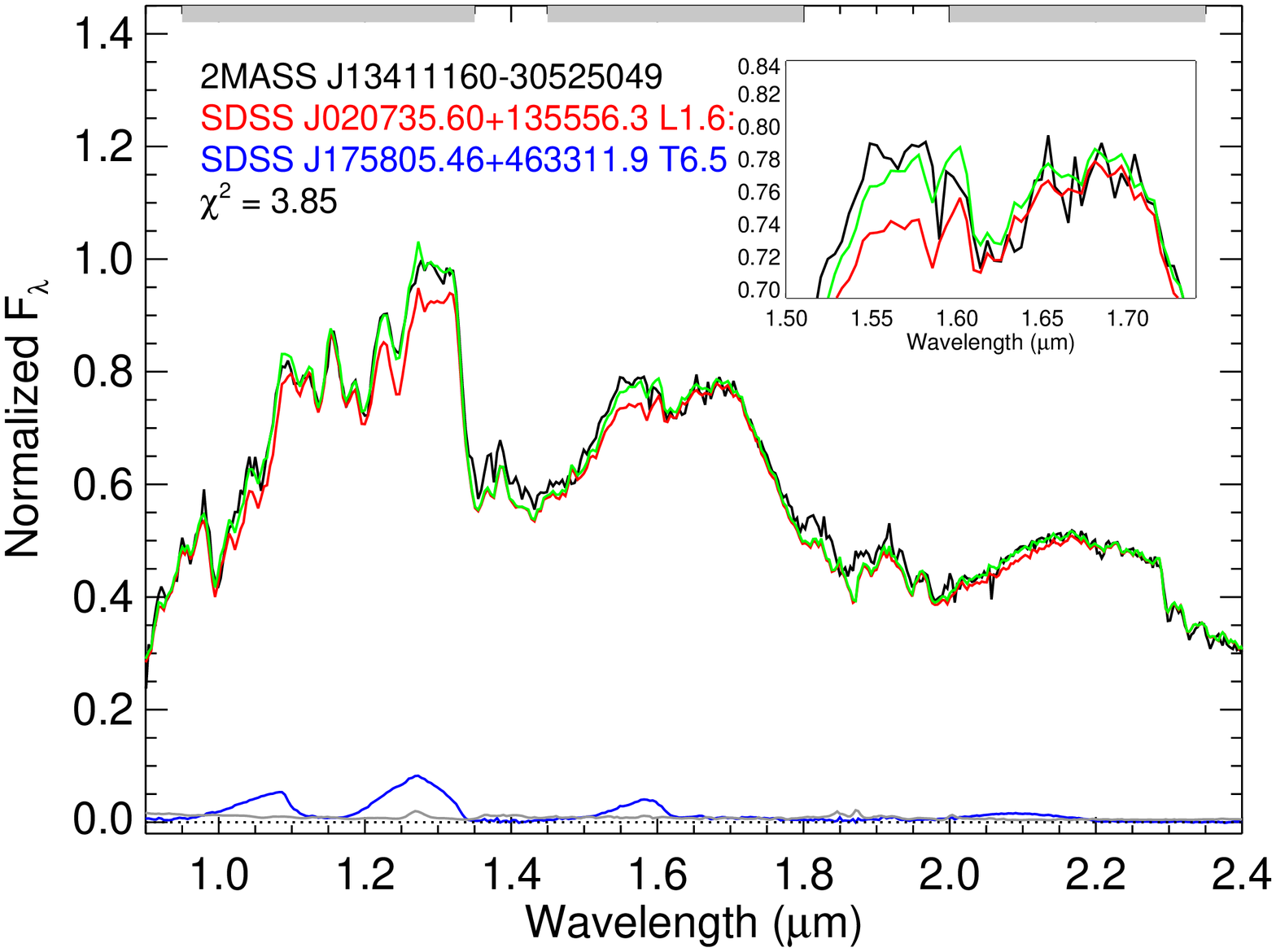}
\includegraphics[scale=0.3]{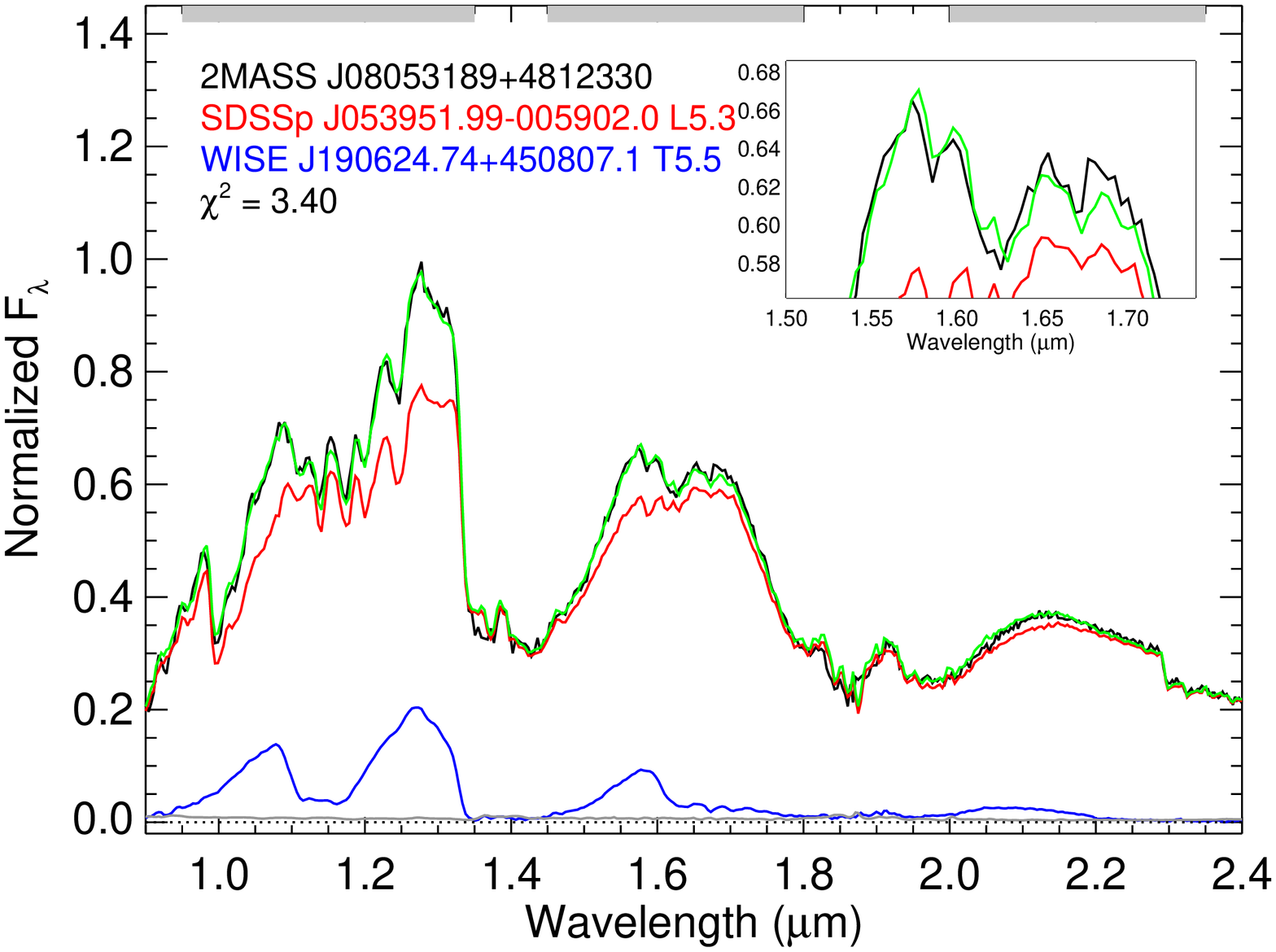}
\caption{Spectral binary fitting for WISE J0720$-$0846, 2MASS J1341$-$3052 and SDSS J0805+4812. The black line is the unresolved spectrum of the source, the red and blue lines correspond to the primary and secondary, respectively, which added together become the best binary template fit (green line).}\label{fig:spexbinaryfit}
\end{figure}

Spectral binaries can be identified by visual inspection, using spectral indices and spectral fitting. The peculiar features that strongly suggest binarity in NIR spectra with late-M/early-L and T components are:
\begin{itemize}
\item Additional flux (as compared to a single object of the same spectral type) in the $J$-band at $1.25~\mu$m, coming from the T dwarf companion;
\item A small dip in the $H$-band at $1.63~\mu$m, resulting from overlapping CH$_4$ (T dwarf) and FeH (M dwarf) absorption features; and 
\item An inflated bump in the $K$-band shortward of $2.2~\mu$m, also coming from the flux from the T dwarf.
\end{itemize}
From a sample of $\sim800$ M7-L7 brown dwarfs near-infrared (NIR) spectra from the SpeX Prism Libraries~\citep{bur14} which excludes poor quality spectra, giants, and optical subdwarfs, we identified 35 candidates through visual inspection and spectral indices. We designed spectral indices that target the features described above, indices previously defined~\citep{bur10, bur06a} colors and spectral type for a total of seventeen parameters. The parameters were compared against each other to select regions where it would be likely to find other binaries guided by four binary benchmarks (See Figure~\ref{fig:indices}). Subsequently, we conducted single and binary template fitting on the 35 candidates, using a chi-squared minimization routine and an F-test to compare the single and binary fits (See Figure~\ref{fig:spexbinaryfit}). Out of the 35 candidates, 18 showed a confidence $>90\%$ that a binary template reproduced the spectra better than a single template. For four sources we deemed their peculiar-looking spectra be due to their blue colors, instead of binarity, so they were rejected. The final count yielded 14 binary candidates.

After performing a Kolmigorov-Smirnov test comparing the separations of a subsample of six confirmed spectral binaries to the rest of the binary sample\footnote{For a fair comparison, both samples were limited to a distance less than $30$pc, and within angular separations of $50-500$mas.}, we find a probability of 25\% that the spectral binary subsample does not belong to the greater binary sample suggesting that the peak may be a selection effect. Given the small number statistics, confirmation of the remaining 43 spectral binary candidates is needed.

\section{Examples of spectral binaries}

\subsection{WISE J072003.20-084651.2}
Originally discovered by~\citet{sch14}, the M9 WISE J$0720-0846$ was identified as a spectral binary of M9.0$\pm$0.5 and T5.0$\pm$0.7 components in Burgasser et al. (2014, submitted) due to its peculiar NIR spectrum. A Keck II/NIRC2 subtracted image in the $H$-band (Figure~\ref{fig:nirc2}) shows a marginally resolved candidate secondary $\Delta H = 4.12\pm0.38$ mag fainter than the primary. This is a very tight binary with angular separation of $0\farcs13$ at a distance of $7.0\pm1.9$pc, meaning that the components in this system are separated by $0.9$AU. Its spectrum shows a strong, broadened and variable H$\alpha$ emission line but lacks the 6708\AA~Li I absorption feature.

\subsection{2MASS J13411160$-$30525049}
\begin{figure}
\centering
\includegraphics[scale=0.04]{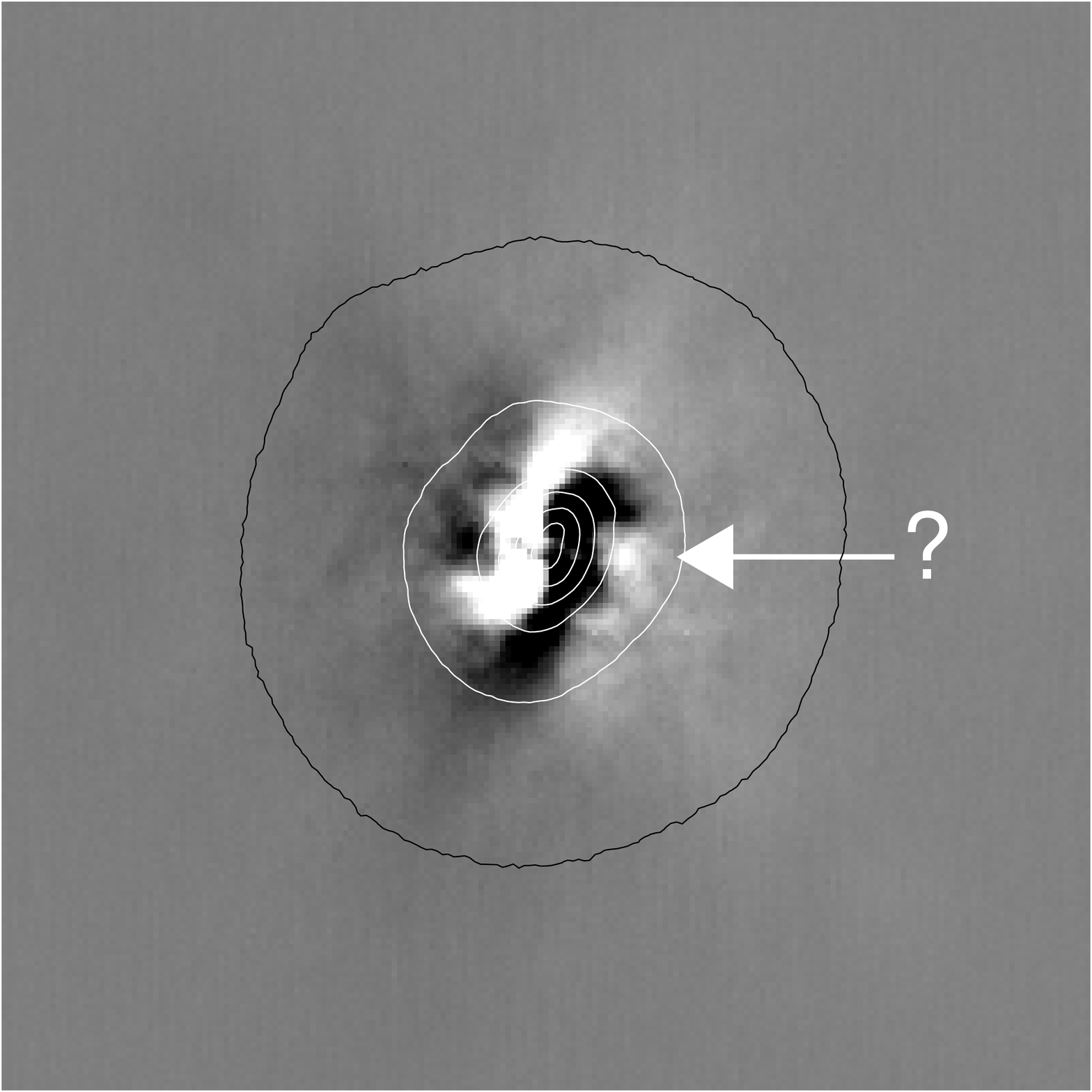}
\includegraphics[scale=0.275]{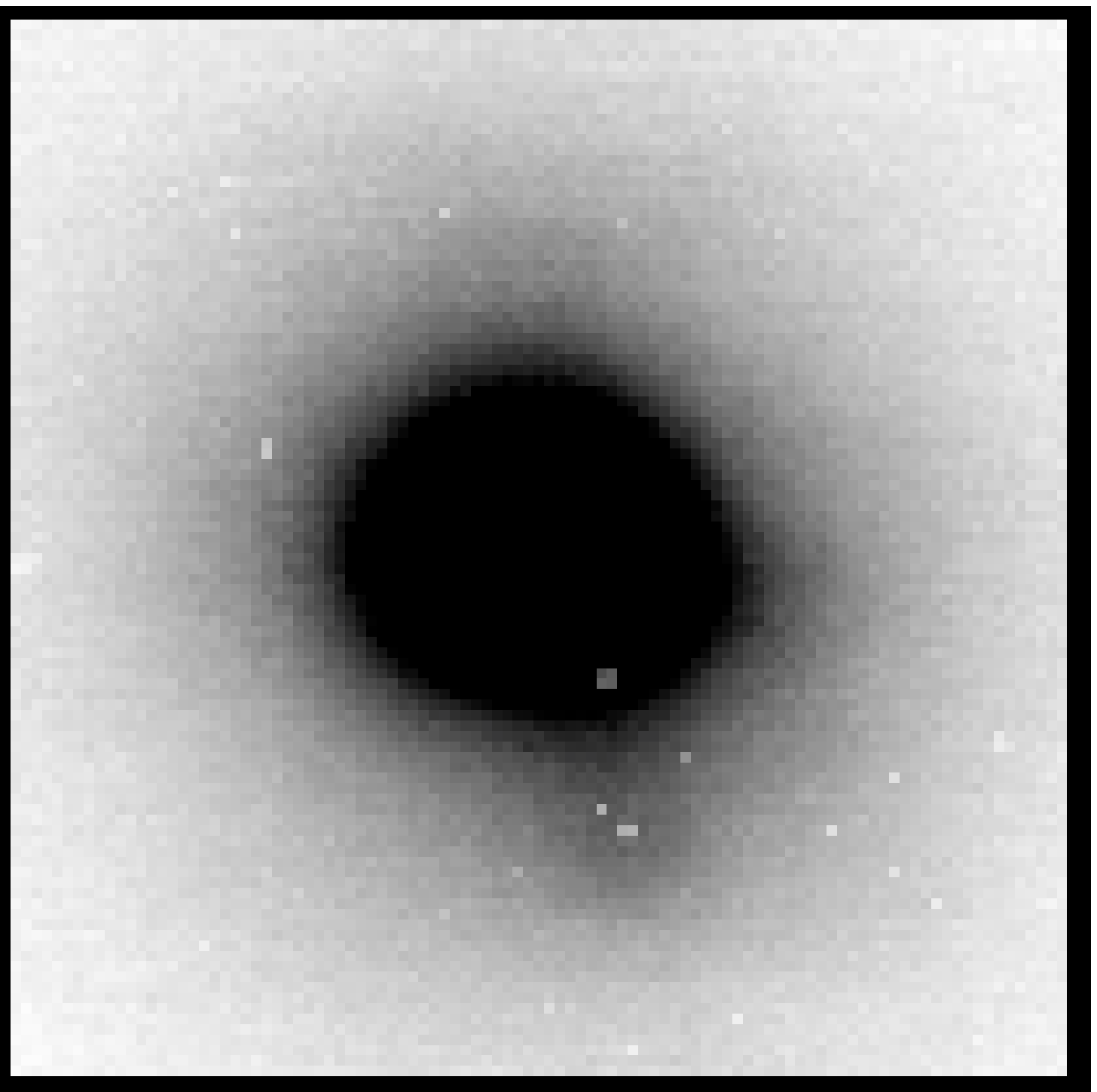}
\caption{Keck II/NIRC2 $H$-band images of WISE J0720-0846 (left), rotated 180 $^\circ$ and subtracted to mimic its own PSF, and 2MASS J1341$-$3052 (right). Figure from Burgasset et al. (2014, submitted).}\label{fig:nirc2}
\end{figure}

2MASS J1341$-$3052 was discovered by~\citet{rei08}, classified as an L3 in the optical and NIR and identified as a spectral binary of L1.0$\pm$0.3 and T6.0$\pm$1.0 components in Bardalez Gagliuffi et al. (submitted). High resolution imaging with Keck II/NIRC2 (Figure~\ref{fig:nirc2}) has been able to resolve the secondary in the $H$-band, $\Delta H = 4.2\pm0.6$ mag fainter than the primary. This binary lies 24$\pm$2 pc away, its angular separation is 0$\farcs$3 thus implying a projected separation of $\sim 8$~AU.

\subsection{SDSS J080531.84+481233.0}
\begin{figure}
\centering
\includegraphics[scale=0.5]{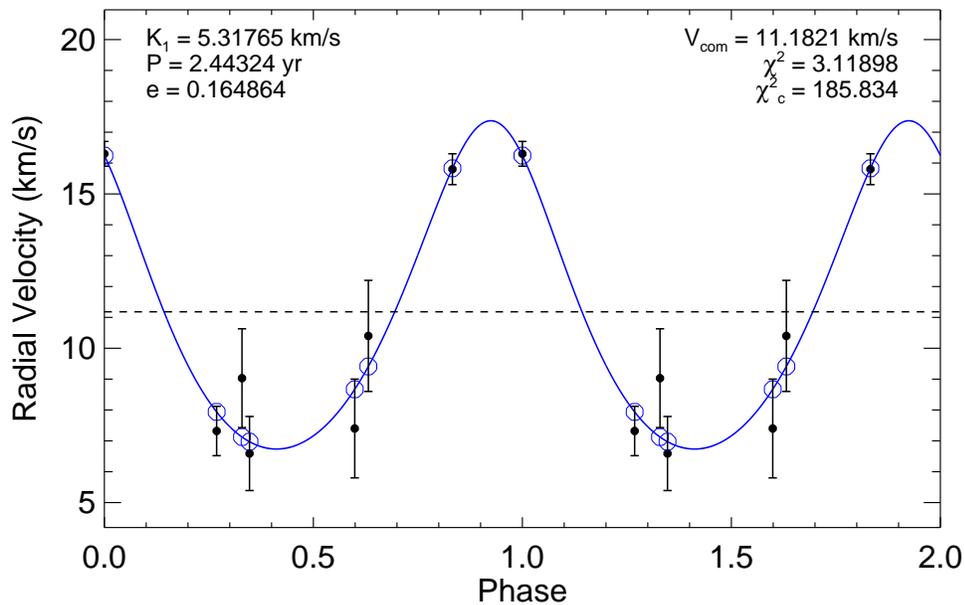}
\caption{Radial velocity curve for SDSS J0805+4812 showing a period of 2.4 years. Figure from Burgasser et al. (in prep).}\label{fig:0805}
\end{figure}

The unusually blue L dwarf SDSS J0805+4812 was identified as a spectral binary by~\citet{bur07} from its discrepant optical and NIR spectral types (L4 and L9.5, respectively) and due to the prominent methane absorption feature in its $H$-band, uncharacteristic of early to mid-L dwarfs. Its peculiar spectrum is best fit with L4.5$\pm$1.0 and T5.0$\pm$0.5 components. Burgasser et al. (in prep) have recently measured a 2.4 year orbit with a radial velocity amplitude of 5 km s$^{-1}$ with Keck II/NIRSPEC (Figure~\ref{fig:0805}).

\acknowledgments{
DCBG acknowledges funding from the IPAC Fellowship from the California Institute of Technology and the Scholarship from the Friends of the International Center at UCSD. This publication makes use of the SpeX Prism Libraries (maintained by Adam Burgasser), the Dwarf Archives Compendium (maintained by Chris Gelino), and the VLM Binaries Archive (maintained by Nick Siegler) based on data obtained with the Infrared Telescope Facility, which is operated by the University of Hawaii under Cooperative Agreement No. NNX-08AE38A with the National Aeronautics and Space Administration, Science Mission Directorate, Planetary Astronomy Program, and at the W.M. Keck Observatory, which is operated as a scientific partnership among the California Institute of Technology, the University of California and the National Aeronautics and Space Administration. The Keck Observatory was made possible by the generous financial supposer of the W.M. Keck Foundation.
}

\end{document}